\documentstyle[aps,preprint,epsf]{revtex}

\begin{document}
\draft 
\title{
Renormalized field theory and particle density profile in driven
diffusive systems with open boundaries
}
\author{H. K. Janssen and K. Oerding}
\address{
Institut f\"ur Theoretische Physik III\\
Heinrich-Heine-Universit\"at\\
D-40225 D\"usseldorf, Germany
}
\date{\today}
\maketitle

\begin{abstract}
We investigate the density profile in a driven diffusive
system caused by a plane particle source perpendicular to the
driving force. Focussing on the case of critical bulk density $\bar{c}$
we use a field theoretic renormalization group approach to calculate the
density $c(z)$ as a function of the distance from the particle source at
first order in $\epsilon=2-d$ ($d$: spatial dimension). For $d=1$ we
find reasonable agreement with the exact solution recently obtained for
the asymmetric exclusion model. Logarithmic corrections to the mean
field profile are computed for $d=2$ with the result $c(z)-\bar{c}
\sim z^{-1} (\ln(z))^{2/3}$ for $z \rightarrow \infty$.
\end{abstract}

\narrowtext

\section{Introduction}

It is well-known that in thermodynamic systems with longe-range
correlations boundaries have a considerable influence on physical
quantities even at macroscopic distances from the surface.
During recent decades this effect has been extensively investigated in
the context of surface critical phenomena (a review is given in
Ref.~\cite{Diehl}). It has been shown that the critical behavior near a
boundary is governed by universal scaling laws with new critical
exponents which cannot be expressed in terms of bulk exponents. The
renormalization group has proved to be a useful method for the
classification of both static~\cite{Diehl,Diet,Diet2,Diet3,Sym} and
dynamic~\cite{DD83,DJ,Diehl2} surface universality classes.

While in equilibrium systems long-range correlations occur if the
thermodynamic parameters approach a critical point, they seem to be the
rule in non-equilibrium systems with conserved dynamics~\cite{GLMS,SZ}.
In this paper we study the diffusion of particles subject to a
driving force in a system with open boundaries. Here the particle
conservation in conjunction with the deviation from detailed balance
leads to long-range correlations and anomalous long-time
behavior~\cite{JS} even at temperatures $T \gg T_{c}$ above the
critical point (for a review see~\cite{SZ}). Especially interesting
from a physical viewpoint are surfaces which act as particle
reservoirs and thus break the conservation law. In the case of
boundaries perpendicular to the driving force which we consider in the
present paper a particle reservoir is necessary to maintain a
non-vanishing steady current.

A simple microscopic realization of a driven diffusive system (DDS) is
a lattice gas~\cite{KLS} with hard core repulsion and nearest neighbor
hopping. In a homogeneous state of density (particles per lattice site)
$c$ the external field (favoring jumps in the positive $z$ direction)
produces a steady particle current $j(c)$. Due to the excluded volume
constraint the current vanishes if every site is occupied by a particle,
i.e. $j(1)=0$, while $j(c) \sim c$ for $c \ll 1$. In a work by
Krug~\cite{JK} on boundary-induced phase transitions the general form
of the density profile has been discussed. It has been shown that the bulk
density $\bar{c}$ satisfies the following maximum current principle:
If the boundary at $z=0$ is in contact with a particle reservoir of
density $c(0)$ and every particle that reaches the boundary at $z=L$
leaves the system, i.e. $c(L)=0$, then the current is maximized in the
sense that 
\begin{equation}
j(\bar{c}) = \max \{j(c) \mid 0 \leq c \leq c(0) \} \label{maxcur}
\end{equation}
for $L \rightarrow \infty$. As a direct consequence of this principle the
bulk density is at the maximum point $c^{\star}$ of the function $j(c)$
and the profile decays only algebraically from boundary- to bulk-value
 if $c(0) \geq c^{\star} \geq c(L)$.

Recently the exact density profile in one dimension has been calculated for
arbitrary boundary conditions $0 \leq c(0), c(L) \leq 1$~\cite{SD,DEHP}.
These works confirm and generalize a large part of the results
obtained in Ref.~\cite{JK}.

Until now no exact solutions have been found for more complicated
problems like driven diffusion in higher dimensional systems, at a
critical point~\cite{JS2}, or in a medium with quenched
disorder~\cite{BJ}. In these cases the field theoretic approach is
useful to obtain systematic approximations for density profiles or
correlation and response functions. In the present paper we use the exact
solution of the one-dimensional asymmetric exclusion model found
in~\cite{SD,DEHP} to test the accuracy of the renormalization-group
improved perturbation theory. We also extend the analysis to
two-dimensional DDS.

In the next section we present the semi-infinite extension of the continuum
model for DDS introduced in Ref.~\cite{JS}. In Sect.~\ref{3} the boundary
conditions are discussed and the density profile is calculated in a mean
field approximation. Having introduced the renormalization factors which are
required to obtain a well-defined renormalized field theory in
Sect.~\ref{4} we calculate the Gaussian fluctuations around the mean field
 profile. We
use the renormalization group to compute the universal scaling function for
the profile at first order in $\epsilon =2-d$ and compare our result with
the exact one-dimensional profile. For $d=2$ we obtain the logarithmic
corrections to the mean field solution. Sect.~\ref{5} contains a discussion
of our findings and an outlook. In Appendix~\ref{A} some technical details
of the calculation of the surface divergencies at one-loop order are
given. In order to compare our perturbative result with the exact solution
we have to take the continuum limit of the exact profile.
This is done in Appendix~\ref{B}.

\section{The Model}

Some time ago, van Beijeren, Kutner, and Spohn~\cite{vBKS}
introduced a continuum model for the diffusion of particles subject to a
driving force. The anomalous long time behavior of this driven diffusive
system (DDS) was studied by Schmittmann and one of us~\cite{JS} by
renormalization group methods. The equation of motion for the particle
density $c({\bf r}, t)$ in this model is given by the continuity equation 
\begin{equation}
\frac{\partial}{\partial t} c({\bf r}, t) + \nabla {\bf j}({\bf r}, t)
= 0 .  \label{1.1}
\end{equation}
Here the current ${\bf j}({\bf r}, t) $ consists of a diffusive
part, a contribution caused by the driving force ${\bf E}$, and a part
${\bf j}_{R}({\bf r}, t)$ which is assumed to summarize the effects of
the fast microscopic degrees of freedom: 
\begin{equation}
{\bf j}({\bf r}, t) = -{\bf D} \nabla c({\bf r}, t) + \kappa\bigl(
c({\bf r}, t)\bigr) {\bf E} + {\bf j}_{R}({\bf r}, t) .  \label{1.2}
\end{equation}
This form of the current can---in principle---be derived from the
microscopic dynamics by a suitable coarse graining. Since the external
field ${\bf E}$ introduces an anisotropy into the system the coarse
graining in general gives rise to an anisotropic matrix of diffusion
constants ${\bf D}$. 

Expanding the conductivity $\kappa(c)$ in the deviation $s=c-\bar{c}$ of
the density from its uniform average (bulk-) density $\bar{c}$, we have 
\begin{equation}
\kappa(c) = \kappa(\bar{c}) + \kappa^{\prime}(\bar{c}) s + \frac{1}{2}
\kappa^{\prime \prime}( \bar{c}) s^{2} +\ldots .
\label{1.3}
\end{equation}
Neglecting higher order terms in this expansion one obtains the Langevin
equation 
\begin{eqnarray}
\partial_{t} s({\bf r}, t) & = & \lambda \bigl( \triangle_{\bot} + \rho
\partial_{\|}^{2}\bigr) s({\bf r}, t) \nonumber \\
&& + \lambda \partial_{\|} \Bigl( f s({\bf r}, t) +\frac{1}{2} g
s({\bf r}, t)^{2}\Bigr) + \zeta({\bf r}, t) , \label{1.4}
\end{eqnarray}
where $f \propto -E \kappa^{\prime}(\bar{c})$ and $g \propto
-E \kappa^{\prime \prime}(\bar{c})$, and the indices $\|$ and $\bot$
distinguish spatial directions parallel (`longitudinal') and perpendicular
(`transverse') to the driving force. The Langevin force $\zeta = -\nabla
{\bf j}_{R}$ is assumed to be Gaussian with zero mean and the
correlations (after a suitable rescaling of $s$) 
\begin{equation}
\langle \zeta({\bf r}, t) \zeta({\bf r}^{\prime}, t^{\prime}) \rangle =
-2 \lambda (\triangle_{\bot} + \sigma \partial_{\|}^{2}) \delta({\bf r} -
{\bf r}^{\prime}) \delta(t - t^{\prime}) .
\label{1.5}
\end{equation}
It was shown in~\cite{JS} that the model defined by
Eqs.~(\ref{1.4},\ref{1.5}) is complete in the renormalization group
sense, i.e.: further contributions to
(\ref{1.4},\ref{1.5}) as well as non-Gaussian and non-Markovian parts
of the Langevin forces are irrelevant for the long-time and large-distance
properties of the system as long as the diffusion constants are
positive. 

For an infinite system the term proportional to the coupling constant
$f$ in the Langevin equation~(\ref{1.4}) can be eliminated by a suitable
Galilean transformation~\cite{JS}. Such a Galilean transformation can of
course not be applied to a system with time independent surfaces
perpendicular to the driving force. However, Krug has
shown~\cite{JK} that in a system with open boundaries in which the
particles are driven from a reservoir of density~$c_{1}$ to a second
reservoir of density $c_{2}$ with $c_{1} > c^{\star} > c_{2}$ the particle
density far in the bulk takes the value $c^{\star}$ which maximizes
the function $\kappa(c)$. In this case we have $f \propto -
\kappa^{\prime}(c^{\star}) = 0$ and $g \propto - \kappa^{\prime
\prime}(c^{\star}) >0$, and the density decreases only algebraically
from the boundary value to the bulk value. Since we are interested in
the behavior of the system near one of the boundaries we may effectively
consider a semi-infinite system with bulk density $\bar{c} = c^{\star}$,
i.e. $f=0$.

Appropriate boundary conditions for diffusive semi-infinite systems have
been derived in~\cite{DJ,Diehl2}. This approach crucially rests on the
assumption of detailed balance. In Ref.~\cite{JS} it was shown that
the dynamics defined by the bulk stochastic equations of
motion~(\ref{1.4},\ref{1.5}) with $f=0$ satisfies this assumption if $w =
\sigma/\rho = 1$. Under renormalization group transformations the
coupling constant $w$ flows to the detailed balance value $w_{\star}=1$
if $d \leq 2$. For $d>2$ the system displays mean field-like behavior.
It should be remarked that for $w\neq 1$ the correlations are
long-ranged~\cite{ZS}. Thus for $d\leq 2$ the variable $(w-1)$ plays the
role of a dangerously irrelevant coupling and its effects have to be
studied.

To set up a renormalized field theory it is convenient to recast the
bulk model in terms of the dynamic
functional~\cite{J,DD,BJW,DP,J1,J2,JS}
\begin{eqnarray}
{\cal J}_{b}[\tilde{s}, s] & = & \int dt \int_{V} d^{d}r \Bigl[\tilde{s}
\partial_{t} s + \lambda \Bigl((\nabla_{\bot}\tilde{s})(\nabla_{\bot}s)
\nonumber \\
&& + \rho (\partial_{\|}\tilde{s})(\partial_{\|}s) + f
(\partial_{\|}\tilde{s}) s + \frac{1}{2} g (\partial_{\|}\tilde{s}) s^{2}
\nonumber \\
&& - (\nabla_{\bot}\tilde{s})^{2} - \sigma
(\partial_{\|}\tilde{s})^{2}\Bigl)\Bigr] , 
\label{1.6}
\end{eqnarray}
where $\tilde{s}$ is a Martin-Siggia-Rose response field~\cite{MSR}. The
range of integration with respect to ${\bf r}$ is the $d$-dimensional
half-space $V = \{ {\bf r} = ({\bf r}_{\bot},z) \mid {\bf r}_{\bot}\in
{\bf R}^{d-1}, z\geq 0 \}$. Correlation and response functions in the
bulk can now be expressed as functional averages with weight $\exp(-{\cal
J}_{b})$. For $w=\sigma /\rho =1$ the functional ${\cal J}_{b}$ can be
written in the detailed balance form~\cite{J1,J2} 
\begin{equation}
{\cal J}_{b}[\tilde{s}, s]=\int dt\int_{V} d^{d}r \Bigl[ \tilde{s} \Bigl(
\partial_{t} s + {\bf R}_{b} \bigl( \frac{\delta {\cal H}}{\delta s} -
\tilde{s}\bigr) \Bigr) \Bigr] , \label{1.7}
\end{equation}
with the reaction kernel 
\begin{equation}
{\bf R}_{b} = \lambda \bigl[ \bigl( \loarrow{\nabla}_{\bot}
\nabla_{\bot} + \rho \loarrow{\partial}_{\|} \partial_{\|}
\bigr) + \frac{g}{3} \bigl( \loarrow{\partial}_{\|} s - s
\partial_{\|} \bigr) \bigl] .  \label{1.8}
\end{equation}
Here $\loarrow{\nabla}$ and $\loarrow{\partial}$ act to the left, while
$\nabla$ and $\partial$ act as usual to the right. The Hamiltonian 
\begin{equation}
{\cal H}[s] = \int_{V} d^{d}r \frac{1}{2} s({\bf r})^{2} ,  \label{1.9}
\end{equation}
defines the purely Gaussian stationary state distribution. Obviously,
${\cal J}_{b}$ now obeys the detailed balance symmetry~\cite{J1,J2} 
\begin{eqnarray}
s(t) & \rightarrow & -s(-t)   \nonumber \\
\tilde{s}(t) & \rightarrow & \tilde{s}(-t) - \left. 
\frac{\delta {\cal H}}{\delta s}\right| _{-t} = \tilde{s}(-t)
-s(-t) .  \label{1.10}
\end{eqnarray}
The last equation implies in particular that 
\begin{equation}
\langle s({\bf r},t) \tilde{s}({\bf r}^{\prime }, t^{\prime}) \rangle =
\Theta(t-t^{\prime}) \langle s({\bf r} ,t) s({\bf r}^{\prime},
t^{\prime}) \rangle . \label{1.11}
\end{equation}

We now turn to the surface contributions ${\cal J}_{1}$ in the full
dynamic functional  ${\cal J}={\cal J}_{b}+{\cal J}_{1}$ arising from
the boundary layer at $z=0$. Locality is assumed for the bulk and the
surface. Therefore ${\cal J}_{1}$ should be written as a
$(d-1)$-dimensional integral over surface fields alone. It is easily seen
that for the Hamiltonian ${\cal H}$ only irrelevant surface terms can be
constructed. Thus ${\cal H}$ has the form~(\ref{1.9}) also in the
semi-infinite case. To retain the detailed balance form~(\ref{1.8}) of
${\cal J}$ we can only modify the reaction kernel~(\ref{1.8}) ${\bf
R}_{b} \rightarrow {\bf R} = {\bf R}_{b} + {\bf R}_{1}$. Such a
modification has to describe the breakdown of particle conservation due
to the reservoir~\cite{Diehl2}. The relevant contributions which respect
the symmetry~(\ref{1.10}) are 
\begin{equation}
{\bf R}_{1} = \lambda \bigl(\tilde{c}+b( s-2\tilde{s}) \bigr)
\delta(z) \label{1.12}
\end{equation}
up to redundant terms. Since the kernel ${\bf R}$ has to be positive
for reasons of stability it follows that $\tilde{c}>0$. This completes our
construction of the dynamic functional in the case of detailed balance.
We eventually get
\widetext
\begin{eqnarray}
\lefteqn{{\cal J}[\tilde{s}, s]} \nonumber \\
& = & \int dt \int d^{d}r \Bigl[ \tilde{s} \Bigl(
\partial_{t} s + {\bf R} \bigl( \frac{\delta {\cal H}}{\delta s} -
\tilde{s} \bigr) \Bigr) \Bigr] \nonumber \\
& = &\int dt \int_{V} d^{d}r \Bigl[ \tilde{s} \partial _{t} s + \lambda
\Bigl( (\nabla_{\bot} \tilde{s}) (\nabla_{\bot} s) + \rho (\partial_{\|}
\tilde{s})(\partial_{\|} s) \Bigr)
+ \frac{1}{2} \lambda g (\partial_{\|} \tilde{s}) s^{2} - \lambda
\Bigl( (\nabla_{\bot} \tilde{s})^{2} + \rho (\partial_{\|}
\tilde{s})^{2}\Bigr) \Bigr]   \nonumber \\
&& + \int dt\int_{\partial V} d^{d-1}r_{\bot} \lambda \Bigl[ \tilde{c}
\tilde{s} (s - \tilde{s}) + b \tilde{s} (s - \tilde{s}) (s-2 \tilde{s})
- \frac{1}{6}g \tilde{s}s^{2} \Bigr] . \label{1.13}
\end{eqnarray}
\narrowtext
We now consider the modifications that we expect if detailed balance
does not hold. First of all, the bulk functional is given by~(\ref{1.6})
with a noise constant $\sigma$ independent of the diffusion constant
$\rho$. In addition, the different surface fields which show up
in~(\ref{1.13}) are now independent, and a boundary source has to be
added to the equation of motion~(\ref{1.4}) leading to a term linear in 
$\tilde{s}_1$. We thus get for the surface dynamic functional
\begin{eqnarray}
{\cal J}_{1}[\tilde{s}, s] & = & \int dt \int_{\partial V}d^{d-1}r_{\bot}
\lambda \Bigl( c \tilde{s} s - \tilde{c} \tilde{s}^2 - \tilde{h}_{1}
\tilde{s} \nonumber \\
&& + \frac{1}{2} g_{a}  \tilde{s} s^2 -  \frac{1}{2} g_{b} \tilde{s}^2
s + \frac{1}{6} g_{c} \tilde{s}^3 \Bigr) .  \label{1.14}
\end{eqnarray}
In Eqs.~(\ref{1.13},\ref{1.14}) we have omitted the surface operators
$s\partial_{n}\tilde{s}$, $\tilde{s}\partial_{n}s$,
$\tilde{s}\partial_{n}\tilde{s}$, $\partial_{n}\tilde{s}$, and
$\partial_{n}^{2} \tilde{s}$ (where $\partial_{n} = \partial_{\|}$
means the normal derivative) since they can be expressed in terms of
the composite fields retained in ${\cal J}_{1}$ (see also below).

We remark that we have always used the prepoint time discretization
in the construction of the dynamic functional~\cite{J1,J2}. This
corresponds to the definition $\Theta(t=0) = 0$ and thus allows to omit
all (measure) terms $\propto \Theta(0)$ in ${\cal J}$.

\section{Equations of motion and mean field profile} \label{3}

Introducing the functional
\begin{eqnarray}
\lefteqn{Z[\tilde{J}, J; \tilde{J}_{1}, J_{1}]} \nonumber \\
& = & \int {\cal D}[\tilde{s}, s] \exp\Bigl( -{\cal J}_{b}[\tilde{s}, s]
- {\cal J}_{1}[\tilde{s}_{s}, s_{s}] \nonumber \\
&& + (\tilde{J}, \tilde{s}) + (J, s) + (\tilde{J}_{1},
\tilde{s}_{s}) + (J_{1}, s_{s}) \Bigr) \label{z}
\end{eqnarray}
we may write correlation and response functions as derivatives of
$Z$ with respect to the bulk sources
$\tilde{J}$, $J$ and the surface sources $\tilde{J}_{1}$, $J_{1}$. In
Eq.~(\ref{z}), $\tilde{s}_{s}$ and $s_{s}$ denote the surface fields and
we have used the abbreviations
\begin{equation}
(\tilde{J}, \tilde{s}) = \int dt \int_{V} d^{d}r \tilde{J} \tilde{s}
\end{equation}
and
\begin{equation}
(J_{1}, s_{s}) = \int dt \int_{\partial V} d^{d-1}r_{\bot} J_{1} s_{s} .
\end{equation}
To obtain the boundary conditions which are determined by the surface
functional ${\cal J}_{1}$ we exploit the invariance of $Z$ with respect
to a shift of the fields $\tilde{s}$
and $s$. This invariance implies the equations of motion
\begin{equation}
\langle \delta {\cal J}_{A} \rangle_{\{\tilde{J}, J;
\tilde{J}_{1}, J_{1}\}} = J_{A}\ \mbox{for $A = \tilde{s}$, $s$,
$\tilde{s}_{s}$, $s_{s}$} , \label{eqmo}
\end{equation}
with the notation $J_{s}=J$ , $J_{\tilde{s}}=\tilde{J}$ , 
$J_{s_{s}}=J_{1}$ and $J_{\tilde{s}_{s}}=\tilde{J}_{1}$ respectively, and
\begin{mathletters}
\begin{eqnarray}
\delta {\cal J}_{\tilde{s}} & = & \partial_{t}s -
\lambda\Bigl((\triangle_{\bot} + \rho \partial_{\|}^{2}) s + f
\partial_{\|}s + \frac{1}{2} g \partial_{\|} s^{2} \nonumber \\
&& - 2 (\triangle_{\bot} + \sigma \partial_{\|}^{2}) \tilde{s}\Bigr) ,
\label{eqma}
\end{eqnarray}
\begin{equation}
\delta {\cal J}_{s} = -\partial_{t} \tilde{s} - \lambda
\Bigl((\triangle_{\bot} + \rho \partial_{\|}^{2}) \tilde{s} - f
\partial_{\|} \tilde{s} - g (\partial_{\|} \tilde{s}) s \Bigr) ,
\label{eqmb}
\end{equation}
\begin{eqnarray}
\delta {\cal J}_{\tilde{s}_{s}} & = & \lambda \Bigl(-\rho\partial_{n} s
+ 2 \sigma \partial_{n} \tilde{s} + (c - f) s_{s} - 2\tilde{c}
\tilde{s}_{s} \nonumber \\
&& + \frac{1}{2} (g_{a}-g) s_{s}^{2} - g_{b} \tilde{s}_{s}
s_{s} + \frac{1}{2} g_{c} \tilde{s}_{s}^{2} - \tilde{h}_{1}\Bigr) ,
\label{eqmc}
\end{eqnarray}
\begin{equation}
\delta {\cal J}_{s_{s}} = \lambda \Bigl(-\rho\partial_{n} \tilde{s}
+ c \tilde{s}_{s} + g_{a} \tilde{s}_{s} s_{s} - \frac{1}{2} g_{b}
\tilde{s}_{s}^{2} \Bigr) . \label{eqmd}
\end{equation}
\end{mathletters}
Equations~(\ref{eqmo},\ref{eqmc},\ref{eqmd}) show that
$\partial_{n} \tilde{s}$ and $\partial_{n}s$ can be expressed in terms
of the surface operators included in ${\cal J}_{b}$. The redundance of
$\tilde{s}_{s} \partial_{n}s$, $s_{s} \partial_{n} \tilde{s}$, and
$\tilde{s}_{s} \partial_{n} \tilde{s}$ follows by differentiating
$\langle \partial_{n} \tilde{s}\rangle$ and $\langle \partial_{n} s
\rangle$ with respect to $\tilde{J}_{1}$ and $J_{1}$, respectively.

Setting $\tilde{J} = J = \tilde{J}_{1} = J_{1} = 0$ in~(\ref{eqmo})
we obtain from~(\ref{eqma}) the equation
\begin{equation}
\partial_{z} \Bigl( \rho \partial_{z} \Phi(z) + f \Phi(z) +
\frac{1}{2} g \left\langle s({\bf r}_{\bot}, z; t)^2 \right\rangle \Bigr)
= 0
\end{equation}
for the steady density profile $\Phi(z) = \langle s({\bf r}_{\bot}, z; t)
\rangle$. Using the boundary condition $\Phi_{\text{bulk}}=0$ which
follows from our definition of $s$ an integration with respect to
$z$ yields
\begin{equation}
\rho \Phi^{\prime}(z) + f \Phi(z) + \frac{1}{2} g \Bigl(C(z) -
C_{\text{bulk}} + \Phi(z)^{2}\Bigr) = 0 , \label{pra}
\end{equation}
where the function $C(z) = \langle (s({\bf r}_{\bot},
z; t) - \Phi(z))^{2} \rangle$ (with the bulk value $C_{\text{bulk}}$)
describes density fluctuations. The boundary condition at
$z=0$ follows from~(\ref{eqmc})~and~(\ref{pra}) as
\begin{equation}
c \Phi_{0} + \frac{1}{2} g_{a} \Bigl(C(0) + \Phi_{0}^{2}\Bigr) -
\frac{1}{2} g C_{\text{bulk}} = \tilde{h}_{1} . \label{prb}
\end{equation}
Note that equal time expectation values such as $\langle s(t)^{n}
\tilde{s}(t)^{m} \rangle$ with $m > 0$ are zero as a consequence of
our prepoint time discretization.
The function $C(z)$ contains an ultraviolet divergence proportional
to $\delta({\bf 0}) = \Lambda_{\bot}^{d-1} \Lambda_{\|}$ (where
$\Lambda_{\bot}$ and $\Lambda_{\|}$ are cut-off wave numbers) which
cancels in Eq.~(\ref{pra}) for $z>0$.

The mean field profile is the solution of Eq.~(\ref{pra}) for vanishing
fluctuations, i.e. $C(z)=0$. In the case $f=0$ which describes the
maximum current phase we obtain
\begin{equation}
\Phi_{\text{mf}}(z) = \Phi_{0} \bigl( 1 + \frac{g}{2\rho} \Phi_{0}
z\bigr)^{-1} . \label{mfprf}
\end{equation}

\section{Renormalization group analysis} \label{4}

\subsection{Surface divergencies}

In the next subsection the density profile will be calculated by an
expansion around the mean field profile~(\ref{mfprf}). Since the
perturbative calculation of Green functions leads to integrals which
are ultraviolet-divergent in the upper critical dimension $d_{c}=2$,
we have to employ a regularization method to obtain a well-defined
perturbation series. We utilize the method of dimensional
regularization which allows us to render all integrals finite up to
poles in $\epsilon = 2-d$. The $\epsilon$-poles are then absorbed
into renormalizations of coupling coefficients and fields (minimal
renormalization). It was shown in Ref.~\cite{JS} that for the bulk
system a renormalization of the coupling coefficients $\rho$ and
$\sigma$ is sufficient to cancel all uv-divergencies.

In general the breaking of translational invariance by a boundary
generates additional divergencies which require additional
renormalizations of surface couplings and fields. (For a review see
Ref.~\cite{Diehl}.) To determine the surface divergencies we first
consider the flat profile $\Phi(z)=0$ which is a steady state of the
system provided that $\Phi_{0}=0$ and $C(z) = C_{\text{bulk}}$ for all
$z>0$ [see Eq.~(\ref{pra})]. [At the stable fixed point with $w =
\sigma/\rho = 1$ the z-independence of $C(z)$ follows immediately from
the the Hamiltonian~(\ref{1.9}).] It will be shown below that this
condition is generally satisfied for $d=1$ and in the high-temperature, strong-field limit in higher dimensions.

The Fourier transform $\hat{G}_{{\bf q}_{\bot}, \omega}(z, z^{\prime})$
of the free (Gaussian) propagator $\langle s({\bf r}_{\bot}, z; t)
\tilde{s}({\bf 0}, z^{\prime}; 0) \rangle$ is the solution of the
differential equation
\begin{equation}
\Bigl( i\omega + \lambda q_{\bot}^{2} - \lambda \rho \partial_{z}^{2}
\Bigr) \hat{G}_{{\bf q}_{\bot}, \omega}(z, z^{\prime}) =
\delta(z-z^{\prime}) \label{propa}
\end{equation}
with the boundary condition
\begin{equation}
\left.\rho \partial_{z^{\prime}} \hat{G}_{{\bf q}_{\bot}, \omega}(z, z^{\prime})\right|_{z^{\prime}=0} = c \hat{G}_{{\bf q}_{\bot},
\omega}(z, 0) . \label{propb}
\end{equation}
This follows in a straightforward way from the terms bilinear in
$\tilde{s}$ and $s$ of the functional ${\cal J}_{b}$ (with $f=0$) and
Eqs.~(\ref{eqmo},\ref{eqmd}). The solution is given by
\begin{eqnarray}
\lefteqn{\hat{G}_{{\bf q}_{\bot}, \omega}(z, z^{\prime})} \\
& = & \frac{1}{2 \lambda \sqrt{\rho} \kappa} \Bigl[ e^{-\kappa
|z-z^{\prime}|/\sqrt{\rho}} + \frac{\kappa - c/\sqrt{\rho}}{\kappa +
c/\sqrt{\rho}} e^{-\kappa (z+z^{\prime})/\sqrt{\rho}} \Bigr] \nonumber
\end{eqnarray}
with
\begin{equation}
\kappa = \Bigl( \frac{i \omega}{\lambda} + q_{\bot}^{2} \Bigr)^{1/2} .
\end{equation}
The parameter $c$ occurring in the surface functional ${\cal J}_{1}$ and
in the propagator describes (for $c>0$) the suppression of density
fluctuations by the particle reservoir at the boundary $z=0$. Since its
momentum dimension is $c \sim \sqrt{\rho} \mu$ the asymptotic scaling
behavior is governed by the fixed point $c_{\star}=\infty$. This means
that up to corrections to scaling the boundary condition~(\ref{propb})
may be replaced by the Dirichlet boundary conditions $\tilde{s}(z=0)=0$
and $s(z=0)=0$.

The Fourier transform of the Gaussian correlator $\langle s({\bf
r}_{\bot}, z; t) s({\bf 0}, z^{\prime}; 0) \rangle$ at the Dirichlet
fixed point follows from~(\ref{1.6}) as
\begin{eqnarray}
\lefteqn{\hat{C}_{{\bf q}_{\bot}, \omega}(z, z^{\prime})} \\
& = & 2 \lambda \int_{0}^{\infty} dy \hat{G}_{{\bf q}_{\bot},
\omega}(z, y) \bigl( q_{\bot}^{2} + \sigma \loarrow{\partial}_{y}
\partial_{y} \bigr) \hat{G}_{-{\bf q}_{\bot}, -\omega}(z^{\prime}, y)
. \nonumber
\end{eqnarray}
Using the differential equation~(\ref{propa}) for the propagator and
the Dirichlet boundary conditions this may be written in the form
\begin{eqnarray}
\lefteqn{\hat{C}_{{\bf q}_{\bot}, \omega}(z, z^{\prime}) = w \bigl[
\hat{G}_{{\bf q}_{\bot}, \omega}(z, z^{\prime}) + \hat{G}_{-{\bf
q}_{\bot}, -\omega}(z^{\prime}, z) \bigr]} \\
&&  + 2 (1-w) \lambda q_{\bot}^{2} \int_{0}^{\infty} dy \hat{G}_{{\bf
q}_{\bot}, \omega}(z, y) \hat{G}_{-{\bf q}_{\bot},
-\omega}(z^{\prime}, y)
, \nonumber
\end{eqnarray}
where $w=\sigma/\rho$. For the calculation of density fluctuations it
is useful to transform the correlator into $({\bf q}_{\bot}, z,
t)$-representation. The result reads
\begin{eqnarray}
\lefteqn{C_{{\bf q}_{\bot}}(z, z^{\prime}; t)} \label{corr0} \\
& = & w G_{{\bf q}_{\bot}}(z, z^{\prime};|t|) + (1-w) \lambda
q_{\bot}^{2} \int_{|t|}^{\infty} G_{{\bf q}_{\bot}}(z,
z^{\prime};t^{\prime}) dt^{\prime} , \nonumber
\end{eqnarray}
where
\begin{eqnarray}
\lefteqn{G_{{\bf q}_{\bot}}(z, z^{\prime};t) = \Theta(t) (4\pi
\lambda \rho t)^{-1/2}} \label{prop0} \\
& & \times e^{-\lambda q_{\bot}^{2} t}
\Bigl[ \exp\Bigl(-\frac{(z-z^{\prime})^{2}}{4\lambda\rho t}\Bigr) -
\exp\Bigl(-\frac{(z+z^{\prime})^{2}}{4\lambda\rho t}\Bigr) \Bigr] .
\nonumber
\end{eqnarray}
is the free progagator.

For $w=1$ Eq.~(\ref{corr0}) reflects the detailed
balance symmetry~(\ref{1.11}). Integrating the correlator~(\ref{corr0}) 
over $q_{\bot}$ we obtain for $t=0$, $z>0$ 
\begin{equation}
C(z) - C_{\text{bulk}} = -(1-w) \frac{(d-1) \Gamma(d/2)}{2 \sqrt{\rho}
(4\pi)^{d/2}} (z/\sqrt{\rho})^{-d} .
\end{equation}
It is thus necessary for the flat profile to be stationary in
dimensions $d>1$ that the condition $w=1$ is satisfied.
Deviations from this stable fixed point generate the long-range
correlations~\cite{GLMS} in the bulk which
are responsible for the $z$-dependence of the function $C(z)$. In this
sense $(w-1)$ is dangerously irrelevant.
There is, however, a microscopic realisation of DDS in which equal-time correlations are absent in spite of the violation of (microscopic)
detailed balance~\cite{Sp}: In the high-temperature, strong-field limit
the particles are non-interacting except for the excluded volume
constraint and cannot jump against the direction of the driving field.
The lack of correlations in this limit can be incorporated into the
field-theoretic description by setting $w=1$ from the outset.
For $d=1$ we can render $w=1$ by a simple rescaling of coupling
coefficients and fields.

We now proceed along the lines taken in the analysis of the surface
critical behavior of equilibrium systems at the ordinary
transition~\cite{Diehl,Diet,DDE}. Since the correlator~(\ref{corr0}) and
the propagator~(\ref{prop0}) vanish for $z^{\prime}=0$ or $z=0$, Green
functions with insertions of the surface fields $\tilde{s}_{s}$ and
$s_{s}$ tend to zero for $c \rightarrow \infty$. A convenient method
to investigate the scaling behavior of quantities that vanish for
$c=\infty$ is the $1/c$-expansion~\cite{Diet,DDE}.

For large $c$ the propagator of the surface field $\tilde{s}_{s}$
behaves as
\begin{equation}
\hat{G}_{{\bf q}_{\bot}, \omega}(z, 0) = c^{-1} \left.\rho
\partial_{z^{\prime}} \hat{G}^{(D)}_{{\bf q}_{\bot}, \omega}(z,
z^{\prime}) \right|_{z^{\prime}=0} + O(c^{-2})
\end{equation}
for $z>0$. In order to derive an analogous relation for the correlator
we have to take into account the surface coupling $\tilde{c}$ in ${\cal
J}_{1}$ which has the same $\mu$-dimension as $c$. By differentiating
the equation of motion~(\ref{eqmo}, \ref{eqmc}) with respect
to $J$ one obtains
\begin{eqnarray}
\left.\rho \partial_{z^{\prime}} \hat{C}_{{\bf q}_{\bot},
\omega}(z, z^{\prime}) \right|_{z^{\prime}=0} & = & 2\sigma \left.
\partial_{z^{\prime}} \hat{G}_{{\bf q}_{\bot}, \omega}(z, z^{\prime})
\right|_{z^{\prime}=0} \\
& + & c \hat{C}_{{\bf q}_{\bot}, \omega}(z, 0) - 2 \tilde{c}
\hat{G}_{{\bf q}_{\bot}, \omega}(z, 0) \nonumber
\end{eqnarray}
and, for $c \rightarrow \infty$,
\begin{eqnarray}
\lefteqn{\hat{C}_{{\bf q}_{\bot}, \omega}(z, 0) = c^{-1} \rho \Bigl[
\left. \partial_{z^{\prime}} \hat{C}^{(D)}_{{\bf q}_{\bot}, \omega}(z,
z^{\prime}) \right|_{z^{\prime}=0}} \nonumber \\
&& - 2\bigl( w - w_{s} \bigr) \left.
\partial_{z^{\prime}} \hat{G}^{(D)}_{{\bf q}_{\bot}, \omega}(z,
z^{\prime}) \right|_{z^{\prime}=0} \Bigr] + O(c^{-2})
\end{eqnarray}
(with $w_{s} = \tilde{c}/c$). Therefore the leading order terms in an
expansion in powers of $c^{-1}$ can be studied in the framework of a
field theory with Dirichlet boundary conditions by the replacements
\begin{mathletters}
\label{repl}
\begin{eqnarray}
\tilde{s}_{s} & \rightarrow & c^{-1} \rho \partial_{n} \tilde{s} \\
s_{s} & \rightarrow & c^{-1} \rho \bigl( \partial_{n} s - 2
(w-w_{s}) \partial_{n} \tilde{s} \bigr)
\end{eqnarray}
\end{mathletters}
in expectation values. To investigate the scaling behavior of Green
functions with insertions of $\tilde{s}_{s}$ and $s_{s}$ we have to
compute the uv-singularities generated by the surface fields on the
r.h.s. of~(\ref{repl}). These singularities require renormalizations
of the form
\begin{mathletters}
\begin{eqnarray}
\rho \partial_{n} \tilde{s} & = & \tilde{Z}_{1}^{1/2} [\rho
\partial_{n} \tilde{s}]_{R} \\
\rho (\partial_{n} s - 2 w \partial_{n} \tilde{s}) & = &
Z_{1}^{1/2} [\rho (\partial_{n} s - 2 w \partial_{n} \tilde{s})]_{R} \\
\tilde{Z}_{1}^{1/2} w_{s} & = & Z_{1}^{1/2} w_{s R} .
\end{eqnarray}
\end{mathletters}
The renormalization factors $\tilde{Z}_{1}$ and $Z_{1}$ are determined
by requiring that the averages
\begin{displaymath}
\langle [\rho \partial_{n} \tilde{s}]_{R} \cdot s(z>0) \rangle
\end{displaymath}
and
\begin{displaymath}
\langle [\rho (\partial_{n} s - 2 w \partial_{n} \tilde{s})]_{R}
\cdot \tilde{s}(z>0) \rangle
\end{displaymath}
be finite for $\epsilon = 0$. Due to causality the term $2 \rho w
\partial_{n} \tilde{s}$ gives no contribution to the second expectation
value but in the Green function $\langle \rho (\partial_{n} s - 2 w
\partial_{n} \tilde{s}) \cdot s(z>0) \rangle$ it is necessary to obtain
a surface field that can be multiplicatively renormalized by $Z_{1}$.

The renormalizations which are necessary to cancel all uv-divergencies
in the translationally invariant bulk theory are given by~\cite{JS}
\begin{equation}
\rho \rightarrow \overcirc{\rho} = Z_{\rho} \rho \qquad
\sigma \rightarrow \overcirc{\sigma} = Z_{\sigma} \sigma \label{zfac}
\end{equation}
with
\begin{mathletters}
\begin{eqnarray}
Z_{\rho} & = & 1 - \frac{u}{8 \epsilon} (3+w) + O(u^{2})
\label{rena} \\
Z_{\sigma} & = & 1 - \frac{u}{16 \epsilon}\bigr( 3 (w + w^{-1})
+ 2\bigl) + O(u^{2}) \label{renb}
\end{eqnarray}
where
\begin{equation}
A_{\epsilon} g^{2}/\rho^{3/2} = u \mu^{\epsilon} . \label{renc}
\end{equation}
\end{mathletters}
In equation~(\ref{renc}), $\mu$ is an external momentum scale and
$A_{\epsilon} = \Gamma(1-\epsilon/2)/((1+\epsilon) (4\pi)^{d/2})$ is a
geometrical factor which has been introduced for convenience. 
The renormalization factors $\tilde{Z}_{1}$ and $Z_{1}$ are calculated
at one-loop order in Appendix~\ref{A} with result
\begin{equation}
\tilde{Z}_{1} = 1 + \frac{w-1}{4} \frac{u}{\epsilon} + O(u^{2})
\qquad Z_{1} = 1 + O(u^{2}).
\end{equation}

\begin{figure}[b]
\vspace*{30mm}
\epsfxsize=420pt
\hspace*{5mm}\epsfbox{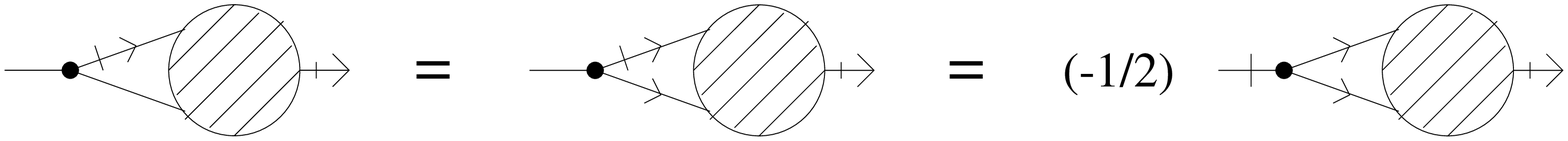}
\vspace*{10mm}
\caption{Perturbative contributions to $\Gamma_{1,1}$.}
\label{feynm}
\end{figure}

We now show that at the stable fixed point $w=1$ the theory is free of
surface singularities, i.e. $\tilde{Z}_{1}=Z_{1}=1$. The idea for the
proof is presented
graphically in Fig.~\ref{feynm}. The hatched bubble represents the sum
of all connected Feynman graphs with a single external
$\tilde{s}$-leg and two $s$-legs. Together with the three-point vertex
representing the bulk coupling $(\lambda g/2)(\partial_{\|}\tilde{s})
s^{2}$ it gives the sum of all perturbative contributions to the
(amputated) Green function $G_{1,1}$. Lines with or without an arrow
denote Dirichlet propagators or correlators while a short line
perpendicular to a propagator/correlator line indicates a derivative
with respect to $z$.
By na{\"\i}ve power counting one expects that the integrations
associated with the first Feynman graph diverge as $\Lambda_{\|}$. Since
the breaking of translational invariance reduces the degree of
divergency by unity~\cite{Diehl} the graph should give rise to
logarithmic surface divergencies. However, Fig.~\ref{feynm} shows that
the graph may be replaced by a different diagram with only logarithmic
bulk singularities (and thus without surface divergencies).

\begin{figure}[b]
\epsfxsize=420pt
\hspace*{5mm}\epsfbox{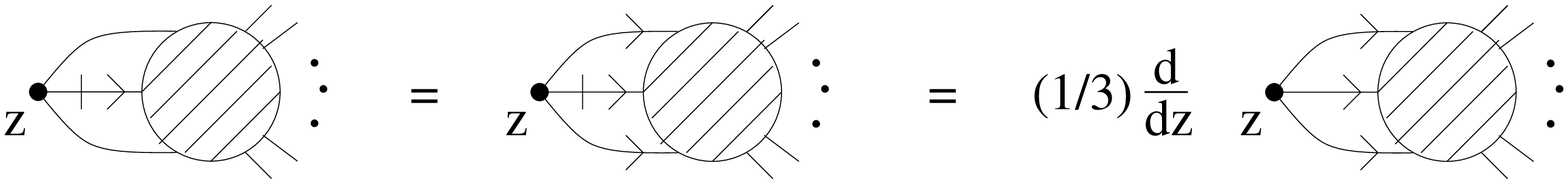}
\vspace*{10mm}
\caption{The vertex is assumed to carry the earliest time argument. In
this case the integration over $z$ vanishes due to the Dirichlet
boundary conditions.}
\label{feyn2}
\end{figure}

The first equality in Fig.~\ref{feynm} is justified if we assume that
the time variable carried by the leftmost vertex is smaller than the
time variables associated with the vertices in the bubble. In this case
we can replace the first correlator by a propagator [Eq.~(\ref{1.11})].
In fact, if we assume that an {\em internal} vertex carries the earliest
time argument the graph vanishes due to the Dirichlet boundary conditions
(see Fig.~\ref{feyn2}). The second equality in Fig.~\ref{feynm} follows
from integration by parts with respect to the $z$-variable of the first
vertex:
\begin{displaymath}
\int_{0}^{\infty} dz s \tilde{s} \partial_{z}\tilde{s} =
\int_{0}^{\infty} dz s \frac{1}{2} \partial_{z} (\tilde{s}^{2}) =
-\frac{1}{2} \int_{0}^{\infty} dz \tilde{s}^{2} \partial_{z} s ,
\end{displaymath}
where we have used the Dirichlet boundary conditions. Since the
$z$-derivative now acts on an external line the degree of divergency is
reduced by unity. The remaining logarithmic bulk divergency is cancelled
by the renormalization of $\rho$.

\subsection{Density profile at one-loop order}

In order to compute the lowest order correction to the mean field
profile we insert the ansatz
\begin{equation}
\Phi(z) = \Phi_{\text{mf}}(z) + \Phi_{1}(z)
\end{equation}
into Eq.~(\ref{pra}) and retain only the terms linear in $\Phi_{1}(z)$.
The resulting differential equation
\begin{equation}
\rho \Phi_{1}^{\prime}(z) + \frac{1}{2} g \bigl( C(z) - C_{\text{bulk}}
+ 2 \Phi_{\text{mf}}(z) \Phi_{1}(z) \bigr) = 0
\end{equation}
has the solution
\begin{eqnarray}
\lefteqn{\Phi_{1}(z)} \nonumber \\
& = & - \frac{g}{2 \rho} \int_{0}^{z} dz^{\prime} \exp\Bigl(
- \frac{g}{\rho} \int_{z^{\prime}}^{z}
\Phi_{\text{mf}}(y) dy \Bigr) \bigl(
C(z^{\prime}) - C_{\text{bulk}} \bigr) \nonumber \\ 
& = & - \frac{g}{2 \rho} \int_{0}^{z} dz^{\prime} \left(
\frac{z^{\prime}+a}{z+a} \right)^{2} \bigl( C(z^{\prime}) -
C_{\text{bulk}} \bigr) , \label{pr1}
\end{eqnarray}
where $a = 2\rho/(g\Phi_{0})$. At one-loop order it is sufficient to
calculate the density fluctuation $C(z)$ in a Gaussian approximation.
Since this function depends on the density profile we have to generalize
the results of the previous subsection to the case $\Phi(z) \neq 0$. For
this purpose we split the field $s({\bf r})$ into its mean value
$\Phi(z)$ and a fluctuating part $\phi({\bf r})$, which leads to the
dynamic functional
\begin{equation}
{\cal J}_{b}[\tilde{\phi}, \Phi + \phi] = {\cal J}_{0}[\tilde{\phi};
\Phi] + {\cal J}_{G}[\tilde{\phi}, \phi; \Phi] + {\cal
J}_{\text{int}}[\tilde{\phi}, \phi] ,
\end{equation}
where
\begin{equation}
{\cal J}_{0}[\tilde{\phi}; \Phi] = \int dt \int_{V} d^{d}r \lambda
(\partial_{\|} \tilde{\phi}) \bigl( \rho\Phi^{\prime}(z) + \frac{1}{2} g
\Phi(z)^{2} \bigr),
\end{equation}
\begin{eqnarray}
\lefteqn{{\cal J}_{G}[\tilde{\phi}, \phi; \Phi]} \nonumber \\
& = & \int dt \int_{V} d^{d}r \Bigl[ \tilde{\phi} \partial_{t}\phi +
\lambda \bigl( (\partial_{\bot} \tilde{\phi})(\partial_{\bot} \phi) +
\rho (\partial_{\|} \tilde{\phi})(\partial_{\|} \phi) \nonumber \\
&& + g \Phi(z) (\partial_{\|} \tilde{\phi}) \phi - (\partial_{\bot}
\tilde{\phi})^{2} - \rho (\partial_{\|} \tilde{\phi})^{2} \bigr)\Bigr] ,
\label{45}
\end{eqnarray}
and
\begin{equation}
{\cal J}_{\text{int}}[\tilde{\phi}, \phi] = \int dt \int_{V} d^{d}r
\frac{1}{2} \lambda g (\partial_{\|} \tilde{\phi}) \phi^{2} .
\end{equation}
(Throughout this subsection we assume $w=1$, i.e. $\sigma = \rho$.)
The Gaussian propagator satisfies the differential equations
\begin{mathletters}
\label{propz}
\begin{eqnarray}
\bigl[ \kappa^{2} - \rho \partial_{z}^{2} - g \partial_{z} \Phi(z)
\bigr] \hat{G}_{{\bf q}_{\bot}, \omega}(z, z^{\prime}) & = &
\frac{1}{\lambda} \delta(z-z^{\prime}) \\
\bigl[ \kappa^{2} - \rho \partial_{z}^{2} + g \Phi(z) \partial_{z}
\bigr] \hat{G}_{{\bf q}_{\bot}, \omega}(z^{\prime}, z) & = &
\frac{1}{\lambda} \delta(z-z^{\prime})
\end{eqnarray}
\end{mathletters}
which follow from the functional ${\cal J}_{G}$~(\ref{45}). The
solution for $\Phi(z) = \Phi_{\text{mf}}(z)$ may be written in the form
\begin{eqnarray}
\lefteqn{\hat{G}_{{\bf q}_{\bot}, \omega}(z, z^{\prime})} \\
& = & \frac{1}{2\lambda \sqrt{\rho} \kappa} \frac{1}{\zeta^{2}} \left[
\frac{f_{+}(\zeta_{0})}{f_{-}(\zeta_{0})} f_{-}(\zeta_{<}) -
f_{+}(\zeta_{<}) \right] f_{-}(\zeta_{>}) , \nonumber
\end{eqnarray}
where

\begin{equation}
f_{\pm}(\zeta) = (1 \mp \zeta) \exp(\pm \zeta)
\end{equation}
and
\begin{mathletters}
\begin{eqnarray}
\zeta & = & \kappa (z+a) /\sqrt{\rho} , \qquad \zeta_{0} = \kappa a
/\sqrt{\rho} , \\
\zeta_{>} & = & \kappa (\max \{z, z^{\prime}\} +a) /\sqrt{\rho} , \\
\zeta_{<} & = & \kappa (\min \{z, z^{\prime}\} +a) /\sqrt{\rho} .
\end{eqnarray}
\end{mathletters}
To obtain the correlator one has to calculate the integral
\begin{eqnarray}
\hat{C}_{{\bf q}_{\bot}, \omega}(z, z^{\prime}) & = & 2 \lambda
\int_{0}^{\infty} dy \hat{G}_{{\bf q}_{\bot}, \omega}(z,
y) \nonumber \\
&& \times \bigl( q_{\bot}^{2} + \rho \loarrow{\partial}_{y}
\partial_{y} \bigr) \hat{G}_{{\bf q}_{\bot}, -\omega}(z^{\prime}, y)
. \label{corrz}
\end{eqnarray}
We use Eqs.~(\ref{propz}) to simplify the integral by eliminating one of
the two $y$-derivatives in~(\ref{corrz}). The result
\begin{eqnarray}
\lefteqn{\hat{C}_{{\bf q}_{\bot}, \omega}(z, z^{\prime}) = \hat{G}_{{\bf
q}_{\bot}, -\omega}(z^{\prime}, z) + \hat{G}_{{\bf q}_{\bot},
\omega}(z, z^{\prime})} \\
&& - \lambda g \int_{0}^{\infty} dy
\Phi_{\text{mf}}(y) \partial_{y} \bigl[
\hat{G}_{{\bf q}_{\bot}, \omega}(z, y) \hat{G}_{{\bf
q}_{\bot}, -\omega}(z^{\prime}, y) \bigr] \nonumber
\end{eqnarray}
can be further simplified by integration by parts since
\begin{equation}
\hat{G}_{{\bf q}_{\bot}, -\omega}(z^{\prime}, y)
\partial_{y} \Phi_{\text{mf}}(y) =
\hat{G}_{{\bf q}_{\bot}, -\omega}(y, z^{\prime})
\partial_{z^{\prime}} \Phi_{\text{mf}}(z^{\prime}) .
\end{equation}
This manipulation leads to
\begin{eqnarray}
\lefteqn{\hat{C}_{{\bf q}_{\bot}, \omega}(z, z^{\prime}) = \hat{G}_{{\bf
q}_{\bot}, -\omega}(z^{\prime}, z) + \hat{G}_{{\bf q}_{\bot},
\omega}(z, z^{\prime})} \\
&& - \frac{2 \lambda \rho}{(z^{\prime}+a)^{2}} \int_{0}^{\infty}
dy \hat{G}_{{\bf q}_{\bot}, \omega}(z, y)
\hat{G}_{{\bf q}_{\bot}, -\omega}(y, z^{\prime}) .
\nonumber
\end{eqnarray}
The function $C(z)$ describing the Gaussian density fluctuations is
given by the integral of $\hat{C}_{{\bf q}_{\bot}, \omega}(z, z)$ over
${\bf q}_{\bot}$ and $\omega$. Using the semi-group property
\begin{eqnarray}
\lefteqn{G_{{\bf q}_{\bot}}(z, z^{\prime}; t_{1}+t_{2})} \nonumber \\
& = & \int_{0}^{\infty} dy G_{{\bf q}_{\bot}}(z,
y; t_{1}) G_{{\bf q}_{\bot}}(y, z^{\prime}; t_{2}) 
\end{eqnarray}
of the propagator in $({\bf q}_{\bot}, z, t)$-representation we obtain
\begin{equation}
C(z) = C_{\text{bulk}} - \frac{\lambda\rho}{(z+a)^{2}} \int_{{\bf
q}_{\bot}} \hat{G}_{{\bf q}_{\bot}, \omega =0}(z, z) , \label{pr2}
\end{equation}
where
\begin{equation}
C_{\text{bulk}} = 2 \int_{{\bf q}_{\bot}, \omega} \Re \bigl[
\hat{G}_{{\bf q}_{\bot}, \omega}(z, z) \bigr] \propto
\Lambda_{\bot}^{d-1} \Lambda_{\|}
\end{equation}
is independent of $z>0$. The straightforward evaluation of the
$q_{\bot}$-integral in~(\ref{pr2}) yields
\begin{eqnarray}
\lefteqn{C(z)-C_{\text{bulk}} =
-\frac{\Gamma(1-\epsilon/2)}{(4\pi)^{d/2}} \frac{2
\rho^{(1-\epsilon)/2}}{(z+a)^{4}} z^{\epsilon} \Bigl[
\frac{(z+a)^{2}}{\epsilon}} \nonumber \\
&& - \frac{z^{2}}{1+\epsilon/2} - \frac{a z}{\Gamma(1-\epsilon)}
\int_{0}^{\infty} dy \frac{y^{-\epsilon}}{1+ay/(2z)} e^{-y} \Bigr] .
\label{pr3}
\end{eqnarray}
To compute the one-loop correction to the profile one has to
insert~(\ref{pr3}) into Eq.~(\ref{pr1}), giving
\begin{equation}
\Phi_{1}(z) = \frac{\Gamma(1-\epsilon/2)}{(4 \pi)^{d/2}} \frac{g
\mu^{-\epsilon} \rho^{-1/2}}{(z+a)^{2}} \Bigl( \frac{z (\mu
z/\sqrt{\rho})^{\epsilon}}{\epsilon (1+\epsilon)} -
F(z; a, \epsilon)\Bigr) \label{pr4} 
\end{equation}
with
\begin{eqnarray}
\lefteqn{F(z; a, \epsilon) = (\mu/\sqrt{\rho})^{\epsilon} \int_{0}^{z}
dz^{\prime} \frac{{z^{\prime}}^{2+\epsilon}}{(z^{\prime} + a)^{2}}} \\
&& \times \Bigl( \frac{1}{1+\epsilon/2} + \frac{2}{\Gamma(1-\epsilon)}
\int_{0}^{\infty} dy \frac{y^{-\epsilon}}{y+2 z^{\prime}/a} e^{-y}
\Bigr) . \nonumber
\end{eqnarray}
In the limit $\epsilon \rightarrow 0$ we have
\begin{eqnarray}
\lefteqn{F(z; a, \epsilon=0) =  z \Bigl( 1+\frac{a}{z+a} \Bigr)} \\
&& -a \Bigl( \gamma + \ln \bigl(\frac{2 z}{a}\bigr) - \frac{z-a}{z+a}
\exp\bigl(\frac{2z}{a}\bigr) {\rm E}_{1}\bigl(\frac{2z}{a}\bigr) \Bigr)
,  \nonumber
\end{eqnarray}
where $\gamma=0.5772\ldots$ is Euler's constant and ${\rm E}_{1}(x)$
denotes the exponential integral~\cite{Ab}.
Eq.~(\ref{pr3}) shows that the perturbation series is not well-defined
for $\epsilon=0$. This uv divergency is cancelled by the renormalization
of the diffusion constant $\rho$ in the mean field profile which leads
to
\begin{eqnarray}
\Phi(z) & = & \frac{2 \rho}{g} \frac{1}{z+a} \Bigl[ 1 +
\frac{u}{2 (z+a)} \Bigl( \frac{z}{\epsilon} \bigl((\mu
z/\sqrt{\rho})^{\epsilon} - 1\bigr) \nonumber \\
&& - (1+\epsilon) F(z; a, \epsilon) \Bigr) \Bigr] + O({\rm
2-loop}) . \label{pr5}
\end{eqnarray}
Since the coupling constant $g$ is
relevant below two dimensions we may not use the perturbation series at
finite order directly to study the asymptotic behavior of the profile.
In the following subsection we improve the one-loop result Eq~(\ref{pr5}) 
by a renormalization group analysis.

\subsection{Scaling}

The renormalization group transformation allows us to map the profile
for large values of $z$ (compared to microscopic length scales) to
length scales which are accessible to perturbative calculations.
Since the bare coupling coefficients are independent of the momentum
scale $\mu$ the profile satisfies (for $w=1$) the renormalization group
equation
\begin{equation}
\bigl[ \mu\partial_{\mu} + \beta(u) \partial_{u} + \zeta(u) (\rho
\partial_{\rho} + a \partial_{a}) \bigr] \Phi(z;u,\rho,a;\mu) = 0 ,
\label{RGE}
\end{equation}
where~\cite{JS}
\begin{equation}
\zeta(u) = \frac{1}{\rho} \left. \mu \frac{d}{d\mu}\right|_{0} \rho
= -\frac{1}{2} u + O(u^{2})
\end{equation}
and
\begin{equation}
\beta(u) = \left. \mu \frac{d}{d\mu}\right|_{0} u = u \bigl(
-\epsilon - \frac{3}{2} \zeta(u) \bigr)
\end{equation}
are defined as derivatives at fixed bare parameters. Eq.~(\ref{RGE})
can be solved by the method of characteristics, with the solution
\begin{equation}
\Phi(z; u, \rho, a; \mu) = \Phi(z; \bar{u}(l), \rho X(l), a X(l); \mu l)
\label{RG2}
\end{equation}
and associated characteristics
\begin{equation}
l \frac{d}{dl} \bar{u}(l) = \beta(\bar{u}(l)) , \qquad l \frac{d}{dl}
\ln X(l) = \zeta(\bar{u}(l)) . \label{flow}
\end{equation}
Using dimensional analysis and Eq.~(\ref{RG2}) it is easy to show that
the profile satisfies the relation
\begin{eqnarray}
\lefteqn{\Phi(z; u, \rho, a; \mu) = (\mu l)^{d/2} (\rho X(l))^{-1/4}}
\label{68} \\
&& \times \Phi\bigl( \mu l z (\rho X(l))^{-1/2}; \bar{u}(l), 1, \mu l a
(X(l)/\rho)^{1/2}; 1 \bigr) . \nonumber
\end{eqnarray}
For small $l$ the scale dependent coupling coefficient $\bar{u}(l)$
tends to the fixed point $u_{\star}=(4/3)\epsilon + O(\epsilon^{2})$ and
the function $X(l)$ is asymptotically proportional to a power of $l$,
namely
\begin{equation}
X(l) \simeq X_{0} l^{-2\eta} , \qquad \eta = -\frac{1}{2}
\zeta(u_{\star}) = \frac{2-d}{3} ,
\end{equation}
where $X_{0}$ is a non-universal scale factor. By choosing the value
\begin{equation}
l = \bigl[ (X_{0} \rho)^{1/2}/(\mu z) \bigr]^{3/(5-d)}
\end{equation}
for the flow parameter we obtain for $z \gg \sqrt{\rho} \mu^{-1}$ the
scaling form
\begin{equation}
\Phi(z, \Phi_{0}) = \Phi_{0} \Xi\bigl( B \Phi_{0}
z^{(d+1)/(5-d)} \bigr) \label{scal}
\end{equation}
with the scaling function 
\begin{equation}
\Xi(x) = x^{-1} \Phi(1; u_{\star}, 1, \alpha x^{-1}; 1)
\end{equation}
(where $\alpha = (4 A_{\epsilon}/u_{\star})^{1/2}$) and the
non-universal constant
\begin{equation}
B = \bigl( (\rho X_{0})^{1/2} \mu^{\epsilon/3}
\bigr)^{-3(d-1)/(2(5-d))} .
\end{equation}
To derive Eq.~(\ref{scal}) we have used the identity
\begin{equation}
A_{\epsilon} g^{2} = u_{\star} (\rho X_{0})^{3/2} \mu^{\epsilon}
\end{equation}
which follows from the flow equations~(\ref{flow}) and the initial
condition~(\ref{renc}) to express $a$ in terms of $\Phi_{0}$, $\rho$,
$u_{\star}$, $X_{0}$, and $\mu$. Note that for $d=1$ there is no
adjustable parameter in Eq.~(\ref{scal}) since $B$ is unity in this
case. The scaling behavior of the one-dimensional profile was first
predicted by Krug~\cite{JK}.

\begin{figure}[b]
\epsfxsize=460pt
\epsfbox{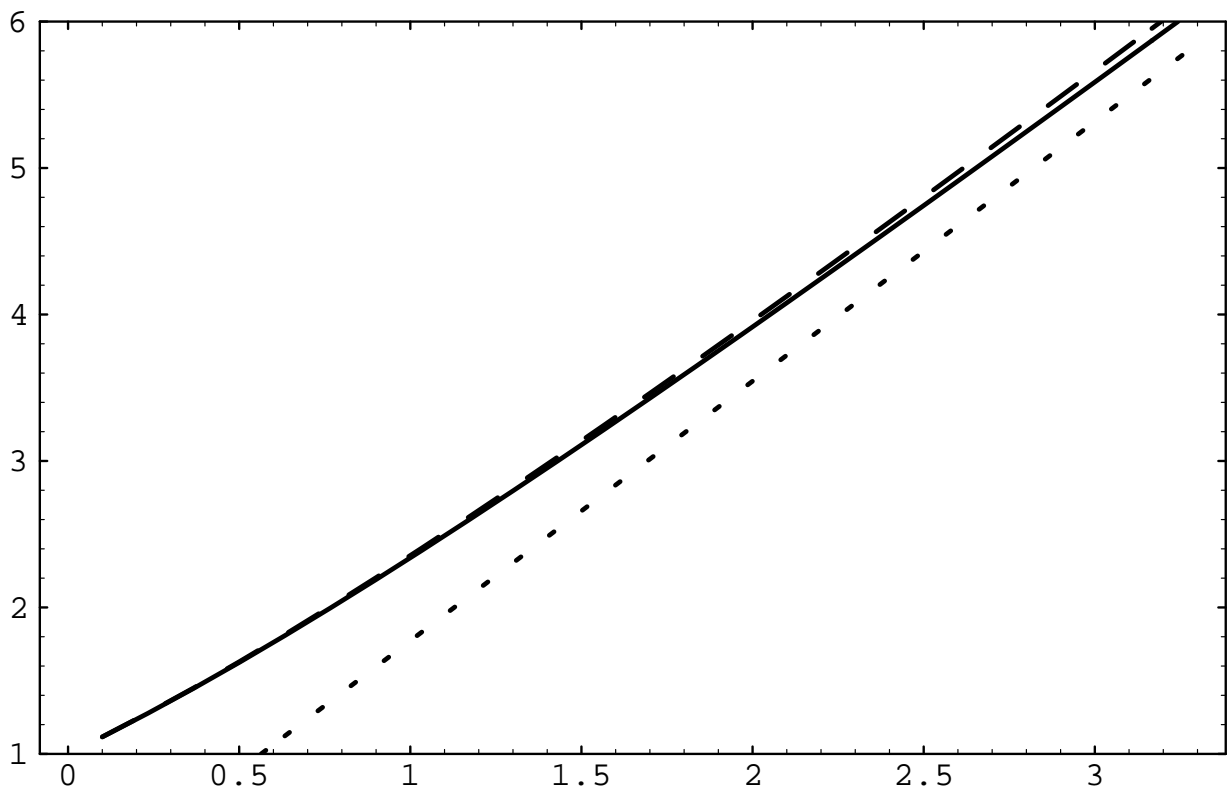}
\vspace*{-35mm}
\caption{Solid curve: scaling function $\Xi_{\text{exact}}(x)^{-1}$
calculated from the exact profile; broken curve: first order in
$\epsilon$; dotted line: asymptotic behavior of
$\Xi_{\text{exact}}(x)^{-1}$ for $x \rightarrow \infty$.}
\label{profile}
\end{figure}

The reciprocal of the scaling function at first order in $\epsilon$ is
given by
\begin{equation}
\Xi(x)^{-1} = 1 + \frac{x}{\alpha} \Bigl(1 + \frac{2 \epsilon}{3}
F(1; \alpha/x, \epsilon=0) \Bigr) + O(\epsilon^{2}) . \label{1l}
\end{equation}
In order to compare this result with the exact solution found in
Refs.~\cite{SD,DEHP} we have to take the continuum limit of the exact
profile. In Appendix~\ref{B} we show that this limit leads to
\begin{equation}
\Xi_{\text{exact}}(x) = \exp(x^{2}) {\rm erfc}(x) ,
\end{equation}
where ${\rm erfc}(x)=\frac{2}{\sqrt{\pi}} \int_{x}^{\infty} dy \exp (-y^{2})$
 denotes the error function~\cite{Ab}. The exact scaling
function and the one-loop approximation~(\ref{1l}) for $\epsilon=1$ are
depicted in Fig.~\ref{profile}. The asymptotic form
\begin{equation}
\Xi_{\text{exact}}(x)^{-1} \simeq \sqrt{\pi} x \qquad \mbox{for $x
\rightarrow \infty$}
\end{equation}
of the scaling function can be compared with the asymptotic behavior
\begin{equation}
\Xi(x)^{-1} \simeq \frac{1}{\alpha} \bigl( 1 + \frac{2}{3} \epsilon +
O(\epsilon^{2}) \bigr) x \approx 1.086 \sqrt{\pi} x
\end{equation}
of the $\epsilon$-expansion at first order for $\epsilon = 1$.
Here and in Fig.~\ref{profile} we have used the geometric prefactor
$\alpha = (4 A_{\epsilon}/u_{\star})^{1/2}$ directly for $\epsilon=1$
(i.e., without expansion in $\epsilon$) since this procedure is
empirically known to give the best numerical results~\cite{JS}.

By a different choice of the flow parameter it is possible to
eliminate the $O(\epsilon)$-contribution to the profile completely.
The corresponding value of $l$ is defined by the equation
[see~(\ref{pr5}, \ref{68})]
\begin{eqnarray}
\lefteqn{\mu l z (\rho X(l))^{-1/2} \ln\bigl( \mu l z (\rho
X(l))^{-1/2} \bigr)} \\
&& - F\Bigl( \mu l z (\rho X(l))^{-1/2}; \mu l a (X(l)/\rho)^{1/2},
\epsilon = 0 \Bigr) = 0 .\nonumber
\end{eqnarray}
However, this modification has only a small effect on the scaling
function $\Xi(x)^{-1}$ at one-loop order.

We conclude this section with the discussion of logarithmic
corrections to the mean field profile in two dimensions. For
$\epsilon=0$, $w=1$ the scale dependent coupling coefficient
$\bar{u}(l)$ is the solution of the differential equation
\begin{equation}
l \frac{d}{dl} \bar{u}(l) = \beta(\bar{u}(l)) = \frac{3}{4}
\bar{u}(l)^{2} + O(\bar{u}(l)^{3})
\end{equation}
with the initial condition $\bar{u}(1)=u$. For $l \rightarrow 0$ we find
the asymptotic expression
\begin{equation}
\bar{u}(l) = \frac{4}{3} \bigl(\ln(1/l)\bigr)^{-1} \Bigl[ 1 + O \Bigl(
\frac{\ln\ln(1/l)}{\ln(1/l)} \Bigr)\Bigr] .
\end{equation}
A straightforward calculation yields the asymptotic behavior of the
characteristic $X(l)$:
\begin{equation}
\ln X(l) = \int_{1}^{l} \frac{dl^{\prime}}{l^{\prime}}
\zeta(\bar{u}(l^{\prime})) \simeq \frac{2}{3} \ln\ln(1/l) + {\rm cst}
\end{equation}
for $l \rightarrow 0$, i.e. $X(l) \sim (\ln(1/l))^{2/3}$. In order to
sum the logarithmic singularities in the perturbation series we set $l =
\sqrt{\rho}/(\mu z)$ in Eq.~(\ref{RG2}) and obtain
\begin{eqnarray}
\Phi(z, \Phi_{0}) & = & \Phi_{0} \Bigl[ 1 + {\rm cst} \times \Phi_{0} z
\bigl( \ln(\mu z/\sqrt{\rho}) \bigr)^{-2/3} \Bigr]^{-1} \label{p2d} \\
& \sim & \frac{(\ln z)^{2/3}}{z} \qquad \mbox{for $z
\rightarrow \infty$.}
\end{eqnarray}
This result is valid on length scales that are large compared to
microscopic distances, i.e. $z \gg \Lambda_{\|}^{-1}$.

\section{Summary and outlook} \label{5}

Until now most analytical studies of driven diffusion deal with closed
systems. To maintain a steady state with a non-vanishing current one
usually imposes periodic boundary conditions in the direction parallel
to the driving force~\cite{SZ}. Since in real physical systems the
particles are fed into the system by a reservoir the effects of
boundaries are of great interest. Using a field theoretic renormalization
group approach we have calculated the density profile in a semi-infinite
DDS with a particle reservoir at the boundary. For $d=1$ our result is
in fair agreement with the exact solution for the totally asymmetric
exclusion model~\cite{SD,DEHP}. Equation~(\ref{p2d}) for the
two-dimensional profile is a new result which may be checked by computer
simulations.

We plan to extend the present work along several lines:

1) We have studied the effect of a particle source (or sink) at $z=0$
assuming a second reservoir to be located at $z = L$ with $ L
\rightarrow \infty$. One would also like to consider the interplay of
the reservoirs for finite $L$~\cite{SD,DEHP}.

2) In a system with quenched disorder and periodic boundary
conditions a particle is subjected to the same random potential after
every passage through the system. In this way the periodicity produces
long-range correlations which make it difficult to compare simulational
results with theoretical predictions which assume uncorrelated
disorder~\cite{BJ}. To circumvent this problem it would be desirable to
consider disordered DDS with open boundary conditions.

3) In the case of non-critical DDS the scaling behavior~(\ref{scal})
of the profile can be described by a single exponent, namely the bulk
exponent $\eta = (2-d)/3$. This is due to the fact that the surface
density $\Phi_{0}$ has the same scaling dimension as the bulk density.
We expect that this is no longer true if the system is at its critical
point. In general the surface critical behavior is governed by new
exponents which cannot be expressed in terms of bulk exponents.

4) Another line of extension would be to consider boundaries parallel
to the driving force.

5) In Sect.~\ref{4} we have shown that for $d>1$ the flat profile
$\Phi(z)=0$ is only stationary in the case of detailed
balance and in the limit of high temperature and strong field.
In the more
general case of a lattice gas at finite temperature the longe-range
correlations in the bulk give rise to an inhomogeneous steady state even
for $\Phi_{0}=\Phi_{\text{bulk}}$. (It was already pointed out by Zia and
Schmittmann~\cite{ZS} that the coupling coefficient $(1-w)$ which is
responsible for the long-range correlations is dangerously irrelevant for
$1 < d \leq 2$.) This effect should lead to a modification of the maximum
current principle [Eq.~(\ref{maxcur})]. Finally, it would be interesting
to study the formation of the stationary density profile starting from
a homogeneous initial state.

\acknowledgments
The authors would like to thank St. Theiss for a critical reading of
the manuscript. This work has been supported in part by the Sonderforschungsbereich 237 [Unordnung und Gro{\ss}e Fluktuationen
(Disorder and Large Fluctuations)] of the Deutsche
Forschungsgemeinschaft.

\appendix

\section{Surface singularities at one-loop order} \label{A}

In Sect.~\ref{4} we have shown that for $w=1$ the perturbation
expansion is free of surface singularities. In this appendix we consider
the general case $w \geq 0$.

The Fourier transform of the surface-bulk response function
\begin{equation}
\chi_{sb}({\bf r}_{\bot}, z; t) = \langle s({\bf r}_{\bot}, z; t)
\cdot \rho \partial_{n} \tilde{s}({\bf 0}, 0) \rangle
\end{equation}
and the bulk-surface resonse function
\begin{equation}
\chi_{bs}({\bf r}_{\bot}, z; t) = \langle \rho \partial_{n}
s({\bf r}_{\bot}, t) \cdot \tilde{s}({\bf 0}, z; 0)  \rangle
\end{equation}
may be written in the form
\widetext
\begin{eqnarray}
\lefteqn{\hat{\chi}_{sb}({\bf q}_{\bot}, \omega, z) = \rho \left.\partial_{z^{\prime}} \hat{G}_{{\bf q}_{\bot}, \omega}(z,
z^{\prime}) \right|_{z^{\prime}=0}} \nonumber \\
&& + \int_{0}^{\infty} dy \int_{0}^{\infty} dy^{\prime} K_{{\bf
q}_{\bot}, \omega}(y, y^{\prime}) \partial_{y} \hat{G}_{{\bf q}_{\bot},
\omega}(z, y) \left.\rho\partial_{z^{\prime}} \hat{G}_{{\bf q}_{\bot},
\omega}(y^{\prime}, z^{\prime}) \right|_{z^{\prime}=0} \label{B1}
\end{eqnarray}
and
\begin{eqnarray}
\lefteqn{\hat{\chi}_{bs}({\bf q}_{\bot}, \omega, z) = \rho \left.\partial_{z^{\prime}} \hat{G}_{{\bf q}_{\bot}, \omega}(z^{\prime},
z) \right|_{z^{\prime}=0}} \nonumber \\
&& + \int_{0}^{\infty} dy \int_{0}^{\infty} dy^{\prime} K_{{\bf
q}_{\bot}, \omega}(y^{\prime}, y) \partial_{y} \hat{G}_{{\bf
q}_{\bot}, \omega}(y, z) \left.\rho\partial_{z^{\prime}} \hat{G}_{{\bf
q}_{\bot}, \omega}(z^{\prime}, y^{\prime}) \right|_{z^{\prime}=0} ,
\label{B2}
\end{eqnarray}
respectively, where $\hat{G}_{{\bf q}_{\bot}, \omega}(z, z^{\prime})$ is
the free propagator~(\ref{prop0}) in Fourier representation. At one-loop
order the kernel $K_{{\bf q}_{\bot}, \omega}(z, z^{\prime})$ is given by
\begin{equation}
K_{{\bf q}_{\bot}, \omega}(z, z^{\prime}) = (\lambda g)^{2} \int_{{\bf
k}_{\bot}, \nu} \bigl[ \partial_{z^{\prime}} \hat{G}_{q_{\bot} -
k_{\bot}, \omega-\nu}(z, z^{\prime})\bigr] \hat{C}_{{\bf k}_{\bot},
\nu}(z, z^{\prime}) + O(g^{4}) . \label{ker}
\end{equation}
\narrowtext
Note that $K_{{\bf q}_{\bot},\omega}(z, z^{\prime})$ has to be treated
as a distribution since the calculation of response functions involves
integrations of $K_{{\bf q}_{\bot},\omega}(z, z^{\prime})$ with respect
to $z$ and $z^{\prime}$.

Before we calculate the singular part of the kernel explicitly it is
helpful to discuss the general form of the divergencies one has to
expect. First, $K_{{\bf q}_{\bot},\omega}(z, z^{\prime})$ has a
non-integrable (for $d=2$) singularity if $z \rightarrow z^{\prime}$.
This divergency occurs in the translationally invariant bulk theory as
well as in the semi-infinite model and has to be cured by a
renormalization of $\rho$. The form of this singularity follows from
dimensional analysis and the isotropy in the directions parallel to the
surface. Since $K_{{\bf q}_{\bot},\omega}(z, z^{\prime}) \sim \lambda
\mu^{2}$ and $z \sim \sqrt{\rho} \mu^{-1}$ the (dimensionally
regularized) bulk singularity contained in $K_{{\bf q}_{\bot},\omega}(z,
z^{\prime})$ is proportional to $\lambda\rho \delta^{\prime}(z -
z^{\prime})$. The breaking of translational invariance by the boundary
generates additional divergencies for $z$, $z^{\prime} \rightarrow
0$~\cite{Diehl}. Upon dimensional regularization they are proportional
to $\lambda\rho \delta(z) \delta(z^{\prime})$.

In order to compute the coefficients of the distributions we apply the
kernel to exponential test functions. At lowest order in $g^{2}$ we have
\widetext
\begin{equation}
\int_{0}^{\infty} dz \int_{0}^{\infty} dz^{\prime} K_{{\bf
q}_{\bot}, \omega}(z, z^{\prime}) e^{-\beta z - \beta^{\prime}
z^{\prime}}
= \frac{\lambda g^{2} \mu^{-\epsilon}}{\epsilon \sqrt{\rho}}
\frac{\Gamma(1 + \epsilon/2)}{(4\pi)^{d/2}} \Bigl[ \frac{3+w}{8}
\frac{\beta^{\prime}}{\beta + \beta^{\prime}} + \frac{w-1}{8}\Bigr] +
O(\epsilon^{0}, g^{4}) .
\end{equation}
The inverse Laplace transform of this expression with respect to
$\beta$ and $\beta^{\prime}$ is given by
\begin{equation}
K_{{\bf q}_{\bot}, \omega}(z, z^{\prime}) = \frac{\lambda
g^{2} \mu^{-\epsilon}}{\epsilon \sqrt{\rho}} \frac{\Gamma(1 +
\epsilon/2)}{(4\pi)^{d/2}} \Bigl[\frac{3+w}{8}
\delta^{\prime}(z^{\prime}|z) + \frac{w-1}{8} \delta(z)
\delta(z^{\prime}) \Bigr] + O(\epsilon^{0}, g^{4}) , \label{sing}
\end{equation}
where we have introduced the definition
\begin{equation}
\int_{0}^{\infty} dz^{\prime} \delta^{\prime}(z^{\prime}|z) f(z^{\prime})
= -f^{\prime}(z).
\end{equation}
(This distribution is equivalent to $\delta^{\prime}(z^{\prime}-z)$ if
we define $\delta(-z) \equiv 0$ for $z \geq 0$.)
It is now straightforward to calculate the singular parts of the
response functions at one-loop order. Inserting~(\ref{sing}) into
Eqs.~(\ref{B1}) and~(\ref{B2}) we find
\begin{equation}
\hat{\chi}_{sb}({\bf q}_{\bot}, \omega, z) = \frac{1}{\lambda}
\Bigl[ 1 + \frac{u}{\epsilon} \Bigl( \frac{w-1}{8} + \frac{3+w}{16}
\frac{\kappa z}{\sqrt{\rho}} \Bigr) + O(\epsilon^{0}, u^{2}) \Bigr]
e^{-\kappa z/\sqrt{\rho}}
\end{equation}
and
\begin{equation}
\hat{\chi}_{bs}({\bf q}_{\bot}, \omega, z) = \frac{1}{\lambda} \Bigl[ 1
+ \frac{u}{\epsilon} \frac{3+w}{16} \frac{\kappa z}{\sqrt{\rho}} +
O(\epsilon^{0}, u^{2}) \Bigr] e^{-\kappa z/\sqrt{\rho}} .
\end{equation}
\narrowtext
To obtain renormalized response functions one has to express the
coupling coefficients $\rho$ and $\sigma$ by their renormalized
counterparts [see Eq.~(\ref{zfac})] and absorb the remaining
$\epsilon$-poles into renormalizations of surface fields, i.e.
\begin{equation}
[\hat{\chi}_{sb}]_{R} = \tilde{Z}_{1}^{-1/2} \hat{\chi}_{sb} , \qquad
[\hat{\chi}_{bs}]_{R} = Z_{1}^{-1/2} \hat{\chi}_{bs}
\end{equation}
with
\begin{equation}
\tilde{Z}_{1} = 1 + \frac{w-1}{4} \frac{u}{\epsilon} + O(u^{2}) ,
\qquad Z_{1} = 1 + O(u^{2}).
\end{equation}
The renormalization factor $Z_{1}$ is unity at every order of the
perturbation series since the primitive surface divergencies in $K_{{\bf
q}_{\bot}, \omega}(z, z^{\prime})$ are proportional to $\delta(z)
\delta(z^{\prime})$. Due to the Dirichlet boundary conditions this
distribution has no effect in Eq.~(\ref{B2}).

\section{The exact profile in the continuum limit} \label{B}

One of the simplest examples of a DDS is the totally
asymmetric exclusion model in one dimension~\cite{DDM}. For this model
with nearest neighbor hopping and open boundary conditions the density
profile has been calculated exactly by Sch\"utz and Domany~\cite{SD} and
Derrida et al.~\cite{DEHP}.

The model is defined on a chain of $N$ sites each of which is either
occupied by a single particle or empty. At each time step one chooses a
random  lattice site $p$, $1 \leq p < N$. If this site is occupied by a
particle ($n_{p}=1$) and its right neighbor site is empty ($n_{p+1}=0$),
the particle will jump from $p$ to $p+1$. In the model with open
boundary conditions particles are injected with rate $\alpha$ at the
left boundary and leave the system with rate $\beta$ at the right
boundary.

For $\alpha, \beta \geq 1/2$ the bulk density takes the critical value
$n_{c}=1/2$ and the system carries the maximum current. Since we wish to
compare the $\epsilon$-expansion of Sect.~\ref{4} with the exact profile
we let $N \rightarrow \infty$ in order to eliminate the influence of the
right boundary. Using the formulas given in the appendix of
Ref.~\cite{SD} we arrive at the particle density
\begin{eqnarray}
\langle n_{p} \rangle & = & \frac{1}{2} + \frac{2}{4^{p}}
\frac{(2p-2)!}{(p-1)!^{2}} \Bigl[ 1 \label{A1} \\
&& - \frac{1}{2p(2\alpha-1)^{2}} F\bigl(1, \frac{3}{2};
p+1; -\frac{4\alpha (1-\alpha)}{(2\alpha-1)^{2}} \bigr)\Bigr] ,
\nonumber
\end{eqnarray}
for $\alpha, \beta \geq 1/2$ and $N \rightarrow
\infty$,
where $F(a, b; c; z)$ is a hypergeometric function~\cite{Ab}.

To compute the scaling function $\Xi_{\text{exact}}(x)$ from the
microscopic profile one has to take an appropriate continuum limit.
In addition to the distance from the boundary the injection rate
$\alpha$ provides the only macroscopic length scale $\xi_{\alpha} =
(2\alpha - 1)^{-2}$ in the model. We therefore measure $p$ in units of
$\xi_{\alpha}$ and let $p, \xi_{\alpha} \rightarrow \infty$
at fixed $x^{2} = p/\xi_{\alpha}$. For this purpose it is convenient to
use the integral representation
\begin{eqnarray}
\lefteqn{F(1, \frac{3}{2}; p+1; -(\xi_{\alpha}-1))} \nonumber \\
& = & p \int_{0}^{1} (1-t)^{p-1} \bigl( 1 + (\xi_{\alpha}-1) t
\bigr)^{-3/2} dt \\
& = & \int_{0}^{p} \exp\bigl( (p-1) \ln(1-\tau/p) \bigr) \nonumber \\
&& \times \bigl( 1 + (\xi_{\alpha}-1) \tau/p \bigr)^{-3/2} d\tau
\end{eqnarray}
to show that
\begin{eqnarray}
\lefteqn{\lim_{p, \xi_{\alpha} \rightarrow \infty} F(1, \frac{3}{2};
p+1; -(\xi_{\alpha}-1))} \nonumber \\
& = & \int_{0}^{\infty} e^{-\tau} (1 + \tau/x^{2})^{-3/2} d\tau .
\end{eqnarray}
For large $p$ the prefactor in Eq.~(\ref{A1}) is given by
\begin{equation}
\frac{2}{4^{p}} \frac{(2p-2)!}{(p-1)!^{2}} = \frac{1}{2 \sqrt{\pi p}}
\bigl( 1 + O(1/p) \bigr) .
\end{equation}
A straightforward calculation now yields in the continuum limit defined
above
\begin{equation}
\langle \sigma_{p} \rangle = \sigma_{0} \exp(\sigma_{0}^{2} p) {\rm
erfc}(\sigma_{0} \sqrt{p}) ,
\end{equation}
where we have expressed the occupation numbers in terms of the spin
variables $\sigma_{p} = 2 n_{p} - 1$, $\sigma_{0} = 2\alpha-1$.

\widetext
\vspace{50mm}

\narrowtext
\vspace*{50mm}
\begin{figure}[b]
\hspace*{5mm}\epsfxsize=420pt
\epsfbox{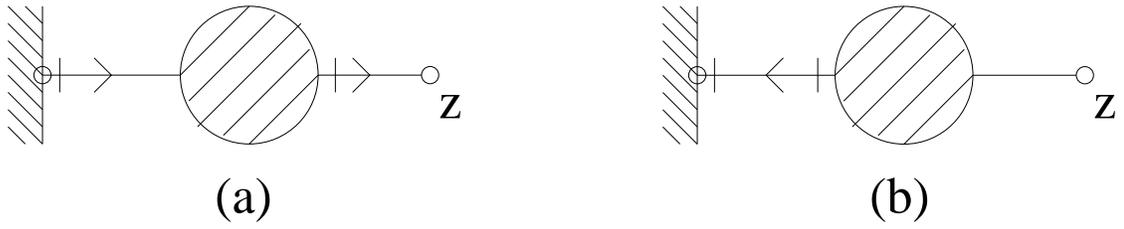}
\vspace*{10mm}
\caption{Perturbative contributions to the (a) surface-bulk and the
(b) bulk-surface response function. Here the hatched bubble represents
the integral kernel~(\protect\ref{ker}).}
\label{1loop}
\end{figure}
\narrowtext

\end{document}